\documentclass[12pt,preprint]{aastex}
\usepackage{multirow}

\shorttitle{MAGNETIC HELICITY AND SOLAR FLARES}
\shortauthors{PARK, CHAE, \& WANG}
\begin{document}

\title{PRODUCTIVITY OF SOLAR FLARES AND MAGNETIC HELICITY INJECTION IN ACTIVE REGIONS}

\author{SUNG-HONG PARK,\altaffilmark{1} JONGCHUL CHAE,\altaffilmark{2} AND HAIMIN WANG\altaffilmark{1}}

\altaffiltext{1}{Space Weather Research Laboratory, New Jersey Institute of Technology, 323 Martin Luther King Boulevard, 101 Tiernan Hall, Newark, NJ 07102, USA; sp295@njit.edu.}
\altaffiltext{2}{Astronomy Program and FPRD, Department of Physics and Astronomy, Seoul National University, Seoul 151-742, Korea.}

\begin{abstract}
The main objective of this study is to better understand how magnetic helicity injection in an active region (AR) is related to the occurrence and intensity of solar flares. We therefore investigate magnetic helicity injection rate and unsigned magnetic flux, as a reference. In total, 378 ARs are analyzed using $SOHO$/MDI magnetograms. The 24 hr averaged helicity injection rate and unsigned magnetic flux are compared with the flare index and the flare-productive probability in next 24 hr following a measurement. In addition, we study the variation of helicity over a span of several days around the times of the 19 flares above M5.0 which occurred in selected strong flare-productive ARs. The major findings of this study are as follows: (1) for a sub-sample of 91 large ARs with unsigned magnetic fluxes in the range from 3 to 5$\times$10$^{22}$ Mx, there is a difference in magnetic helicity injection rate between flaring ARs and non-flaring ARs by a factor of 2; (2) the $GOES$ C-flare-productive probability as a function of helicity injection displays a sharp boundary between flare-productive ARs and flare-quiet ones; (3) the history of helicity injection before all the 19 major flares displayed a common characteristic: a significant helicity accumulation of (3--45)$\times$10$^{42}$ Mx$^2$ during a phase of monotonically increasing helicity over 0.5--2 days. Our results support the notion that helicity injection is important in flares, but it is not effective to use it alone for the purpose of flare forecast. It is necessary to find a way to better characterize the time history of helicity injection as well as its spatial distribution inside ARs.
\end{abstract}

\keywords{Sun: flares---Sun: magnetic topology---Sun: photosphere}

\section{INTRODUCTION}
A solar flare is a sudden, rapid, and intense brightening usually in a magnetically complicated solar active region (AR). Solar flares produce high energy particles, radiation, and erupting magnetic structures that are related to geomagnetic storms. Their strong electromagnetic radiations from radio waves to gamma-rays have direct effect on cell phones and the global positioning system and heat up the terrestrial atmosphere within minutes so that satellites drop into lower orbits (Schwenn 2006). Enormous economic and commercial losses can be caused by these effects (Baker 2004). Therefore there have been significant efforts to develop flare forecasting systems as part of the space weather service. It is generally thought that flare-productive ARs exhibit complex and non-potential magnetic structures related to the stored magnetic energy to power flares. For this reason, many studies of relationship between the solar flare and photospheric magnetic field properties have been carried out since the flare was first observed and recorded by Carrington (1859) and Hodgson (1859). Some examples include the unbalanced changes in the photospheric line-of-sight magnetic field (Cameron \& Sammis 1999; Spirock et al. 2002; Wang et al. 2002); rapid changes of the sunspot structure associated with a substantial fraction of flares (Liu et al. 2005; Deng et al. 2005; Wang et al. 2004a, 2005; Chen et al. 2007); the magnetic shear angle evolution (Hagyard et al. 1984; Hagyard \& Rabin 1986; Sivaraman et al. 1992; Schmieder et al. 1994; Wang et al. 1994, 2004a); the horizontal gradient of longitudinal magnetic fields (Zirin \& Wang 1993; Zhang et al. 1994; Tian et al. 2002); electric current (Canfield et al. 1993; Lin et al. 1993); magnetic helicity injection (Moon et al. 2002a, 2000b; Sakurai \& Hagino 2003; Yokoyama et al. 2003; Park et al. 2008). Based on the above mentioned studies, current flare forecasting models are moving toward multiple-magnetic parameter-based approaches (Leka \& Barnes 2003a, 2003b; Li et al 2008) from sunspot-morphological evolution-based approaches (McIntosh 1990; Gallagher et al. 2002).

Magnetic helicity is a measure of twists, kinks, and inter-linkages of magnetic field lines (Berger \& Field 1984; Pevtsov 2008) and is a useful and important parameter to indicate topology and non-potentiality of a magnetic field system. Naturally magnetic helicity studies have been conducted to the energy buildup and instability leading to solar eruptions and coronal mass ejections (e.g., Rust 2001; Kusano et al. 2004; Phillips et al. 2005; Fan 2009). There were a number of studies related to a rapid magnetic helicity change as an impending condition or a trigger for solar flares (e.g. Moon et al. 2002a, 2002b). LaBonte et al. (2007) surveyed magnetic helicity injection in 48 X-flaring ARs and 345 non-X-class flaring regions, and found that a necessary condition for the occurrence of an X-class flare is that the peak helicity flux has a magnitude $\textgreater$ 6$\times$10$^{36}$ Mx$^2$ s$^{-1}$. Park et al. (2008) found that a substantial amount of helicity is accumulated before the flare in all the 11 X-class flare events, and suggested a warning sign of flares can be given by the presence of a phase of monotonically increasing helicity. Motivated by these results we explore the feasibility of using magnetic helicity to build a flare forecasting system utilizing a data sample covering almost one solar cycle.

\section{DATA AND ANALYSIS}
We use a set of the full-disk 96 minute MDI (Scherrer et al. 1995) magnetogram data to determine 24 hr profiles of magnetic helicity injection rate and unsigned magnetic flux of an AR. Note that the newly calibrated level 1.8.2 MDI magnetograms are employed in this paper. The level 1.8.2 MDI data have been available since 2008 December 24, and its magnetic field value on average over the solar disk increased by a factor of $\sim$1.6 compared to that of the previous level 1.8 data (Tran et al. 2005; Ulrich et al. 2009). A total of 378 ARs were selected during the time period from mid-1996 to 2006, almost the entire duration of Solar Cycle 23. Normally each data set corresponding to a given AR has around 15 MDI magnetograms covering 24 hr during its disk passage. To reduce the effect of the geometrical projection in calculation of the normal component of the magnetic field, we consider the start time of each data set as the time when the corresponding AR appears or rotates to a position within 0.6 of a solar radius from the apparent disk center.

The unsigned magnetic flux, $\Phi$, of the entire area of a given AR is defined by
\begin{equation}
\Phi = \int_s \left|B_n\right| \, dS,
\end{equation}
where $B_n$ is the normal component of the magnetic field; $dS$ is the surface integral element and the integration is over the entire area of the target AR. We approximately determine $B_n$ from the line-of-sight magnetic field, $B_l$, in the MDI magnetograms, assuming that the magnetic field on the solar photosphere is normal to the solar surface, i.e., $B_l= B_n \cos\psi$ where $\psi$ is the heliocentric angle of the point of interest.

We use a simplified expression for the helicity injection rate, $\dot{H}$, of an entire AR given by Chae (2001):
\begin{equation}
\dot{H} = -\int_s 2 \, ({\textit{\textbf{A}}_{p}} \cdot {\textit{\textbf{v}}_{\mathrm{LCT}}}) \, B_n \, dS,
\end{equation}
where $B_n$ is the normal field component, $\textit{\textbf{A}}$$_{p}$ is the vector potential of the potential field, $\textit{\textbf{v}}$$_{\mathrm{LCT}}$ represents the apparent horizontal motion of field lines derived by local correlation tracking (LCT), and $dS$ is the surface integral element over the entire AR area. We determine the quantities in Equation (2) following the procedure described in Chae \& Jeong (2005). $\textit{\textbf{A}}$$_{p}$ is calculated from $B_n$ by using the fast fourier transform method. $\textit{\textbf{v}}$$_{\mathrm{LCT}}$ is calculated using the LCT technique (November \& Simon 1988). For LCT, we align all magnetograms to the first image of the data set after correcting the differential rotation. We set the FWHM of the apodizing window function to 10{\arcsec} and the time interval between two frames to 96 minutes, and performed LCT for all pixels with an absolute flux density greater than 5 G. Only the windows with calculated cross-correlation coefficient above 0.9 are included in constructing velocity maps. After the helicity injection rate is determined as a function of time, we integrate it with respect to time to determine the helicity change over a given day, $\Delta H$.

For each of the 378 AR data sets, we estimate the uncertainty of the helicity injection rate corresponding to measurement uncertainty ($\sim$20 G) of MDI magnetograms. First, we add pseudo-random noise to each magnetogram. The noises have normal distribution with the standard deviation of 20 G. Then we calculate the helicity injection rate calculation described in Equation (2). The same process is repeated 10 times for the same AR domain with different sets of errors to calculate the standard deviations of $\dot{H}$. Finally, we consider the average of the standard deviations during the entire 24 hr period as the uncertainty of the helicity calculation for each AR data set. We found that the average uncertainty of $\dot{H}$ is around 5\% so that we believe it does not significantly affect our helicity calculation and conclusion of the study.

From the time profiles of $\dot{H}$ and $\Phi$, we define the two average parameters to investigate their feasibility for flare forecasting as follows. The first parameter is the absolute value of the average helicity injection rate, $|$$<$$\dot{H}$$>$$|$, which indicates the average amount of injected helicity per unit time to an entire AR defined by
\begin{equation}
|\!\!<\!\dot{H}\!>\!\!|\, = \, \frac{\sum_{t_0}^{t_1}|\dot{H}|}{N},
\end{equation}
where $t_0$ is the start of each data set under investigation, $t_1$ is 24 hr after $t_0$, and $N$ is the total number of MDI magnetograms in each data set during the time period, $\Delta t$, between $t_0$ and $t_1$. As for the second parameter, we use the average unsigned magnetic flux, $<$$\Phi$$>$, with
\begin{equation}
<\!\Phi\!> \, = \,  \frac{\sum_{t_0}^{t_1}\Phi}{N},
\end{equation}
where $t_0$, $t_1$, and $N$ are the same as defined in Equation (4). These parameters were studied by many authors before we use it as a reference. Please refer to the other two helicity parameters which are the maximum values of the data sets of absolute helicity injection rate, $|\dot{H}|$, and absolute helicity accumulation, $|\Delta H|$. They show a similar correlation result with flare productivity as $|$$<$$\dot{H}$$>$$|$ so that in this paper we only represent $|$$<$$\dot{H}$$>$$|$ as a helicity parameter.

To investigate a relationship between these two parameters of an AR and flares occurred in the region for the following day of the parameters' measurement, we use the flare index, $F_{idx}$, which represents each AR's average daily $GOES$ soft X-ray peak flux. $F_{idx}$ was first introduced by Antalova (1996) and was later applied by Abramenko (2005):
\begin{equation}
F_{idx} =  \left(100S^{(X)}+10S^{(M)}+1.0S^{(C)}+0.1S^{(B)}\right)/\tau,
\end{equation}
where $\tau$ is the time interval (measured in days) and $S^{(i)}$ is the sum of $GOES$ flare significants in the $i$th $GOES$ class over $\tau$. We calculate this flare index for each AR, and $\tau$ is selected to be 1 to evaluate the flare productivity of a given region for the time window of the next day following $\Delta t$. Note that in the $GOES$ soft X-ray flare catalog, there are some X-ray events of which locations (indicated as NOAA AR numbers) are unknown so that it might affect our results; however, these are typically weaker events. Wheatland (2001) reported that of the C-class flares in the catalog during the period 1981--1999, 61.5\% are identified with an AR, while of the M- and X-class flares the fractions are 82\% and 94\%, respectively.

\section{RESULT}
In Figure 1, we present 24 hr profiles of $\dot{H}$ and $\Phi$ for all the 378 ARs with three different groups classified by the flare index ranges which are $F_{idx}$$\ge$10 (51 samples, left column), 1$\le$$F_{idx}$$<$10 (74 samples, middle column), and $F_{idx}$$<$1 (253 samples, right column). Note that $F_{idx}$ values of 1, 10, and 100 are equivalent to the specific flare productivity of one C 1.0, M 1.0, and X 1.0 flare per day, respectively. The dotted lines show the average of the maximum values for $|\dot{H}|$ (top panels) and $\Phi$ (bottom panels) of the samples in each panel. As we anticipated, there is a general trend that the larger $F_{idx}$ an AR has, the larger values of helicity injection rate and unsigned magnetic flux it represents. This trend is more evident in the case of $\dot{H}$; the average value (46$\times$10$^{40}$ Mx$^2$ hr$^{-1}$) of the samples for $F_{idx}$$\ge$10 is almost twice greater than that (25$\times$10$^{40}$ Mx$^2$ hr$^{-1}$) of the samples for 1$\le$$F_{idx}$$<$10 and about 4.5 times greater than that (10$\times$10$^{40}$ Mx$^2$ hr$^{-1}$) of the samples for $F_{idx}$$<$1. For the ARs having the large $F_{idx}$, we found that although the magnetic flux does not change too much in time, the helicity, however, accumulates significantly and consistently. This finding agrees with our previous results that 11 X-class flares are preceded by a monotonically significant helicity accumulation, 10$^{42}$--10$^{43}$ Mx$^2$ over a period of half to a few days (Park et al. 2008).

To examine how $\dot{H}$ and $\Phi$ are related with flare productivity, we studied the two average parameters ($|$$<$$\dot{H}$$>$$|$ and $<$$\Phi$$>$) in more detail for the 378 ARs (see Table 1 for the statistical properties of the parameters) and compare them with $F_{idx}$ taken as the proxy for the flare productivity in the next day following the measurement of the parameters. By only considering the 153 samples with non-zero flare index, in Figure 2, we plot, as cross symbols, $F_{idx}$ versus $|$$<$$\dot{H}$$>$$|$ and $<$$\Phi$$>$ in a logarithmic scale. The solid and dotted lines show the least-squares linear fit and its standard deviation to the data points. The correlation coefficients (CCs) and the power law index of the linear fits are also given in each panel. While the data points are widely scattered, there is a moderate correlation between the parameters and the flare index with a tendency that the larger the parameters the larger the flare index. CCs of $F_{idx}$ versus $|$$<$$\dot{H}$$>$$|$ and $<$$\Phi$$>$ are 0.42 and 0.43, respectively. For the other 225 samples with zero flare index, in Figure 2, we marked them as square symbols using $F_{idx}$ = 0.05 for the plotting purpose only. Note that the zero flare index samples are excluded from the linear fit. In the rescaled range of each parameter (the maximum value of the samples is considered as 1 and the minimum as 0), most of the data samples of the helicity parameter are distributed in the range less than 0.2 with flare indexes near zero values, and a few samples are scattered in the range greater than 0.2 with relatively high flare indexes. Instead, $<$$\Phi$$>$ has well-distributed data samples. This difference would make it easier to define a critical value for the helicity parameter to forecast flare-active or flare-quiet conditions.

Furthermore, we are interested in understanding why some of the samples with the large parameters do not produce major flares. Therefore, two groups of samples were selected and their average values for each of the two parameters were compared. The first group contains the samples with $F_{idx}$$\ge$1 and the relatively large parameters (hereafter, $flaring$ group), and the samples in the second group are picked out from the same range of the parameters as those in the $flaring$ group but produced no flares (hereafter, $non$-$flaring$ group). For the comparison, we consider the $flaring$ and $non$-$flaring$ groups for the parameters of $|$$<$$\dot{H}$$>$$|$ and $<$$\Phi$$>$. The samples in the $flaring$ and $non$-$flaring$ groups are marked with the gray boxes in each panel of Figure 2(a) and 2(b). First, for a sub-sample of 118 ARs with large $|$$<$$\dot{H}$$>$$|$ in the range from 2 to 10$\times$10$^{40}$ Mx$^2$ hr$^{-1}$ (Figure 2a), the $flaring$ group has the average unsigned magnetic flux of 3.7$\times$10$^{22}$ Mx, greater than that of the $non$-$flaring$ group, 3$\times$10$^{22}$ Mx. This difference is not significant. However, for a sub-sample of 91 ARs in the range, (3--5)$\times$10$^{22}$ Mx, of large $<$$\Phi$$>$ (Figure 2b), the $flaring$ group has $|$$<$$\dot{H}$$>$$|$ about twice greater than that of the $non$-$flaring$ group. This indicates that in an AR with large flux, a large amount of consistent helicity injection is essential to the occurrence of flares. Please refer to Table 2 for the detailed values of these comparisons.

In Figure 3, we plot $<$$\Phi$$>$ versus $|$$<$$\dot{H}$$>$$|$ in a logarithmic scale for the 378 samples with $F_{idx}$$<$0.1 as plus symbols, 0.1$<$$F_{idx}$$<$10 as triangle, and $F_{idx}$$>$10 as square. $F_{idx}$ is derived for the three different time windows of the first day (Figure 3, top) following $\Delta t$, the second day (Figure 3, middle), and the third day (Figure 3, bottom). In each plot, four sections are determined by the vertical and horizontal dashed lines marking the median values of both the parameters for the samples. By only considering the samples in each of the four sections, we calculate the probability of flare occurrence classified by two groups of criteria $F_{idx}$$\ge$10 and $F_{idx}$$\ge$0.1, and marked as black and gray colored numbers in each section. We found that the flaring probability in the upper right section is not only much greater than those in the upper left and lower right sections but also the probability in the lower right section is always greater than that in the upper left section except only one case for the flare criterion of $F_{idx}$$\ge$0.1 in the second day; especially for the cases of the flare criterion of $F_{idx}$$\ge$10, the probability in the lower right section is almost 2--5 times greater than that in the upper left section. This remarkable thing indicates that magnetic helicity injection will contribute extra weight to improve the flare prediction based on the total unsigned magnetic flux. Another finding is that the flaring probability in the upper right section shows its maximum at the first day time window and it keeps going down for the second and third days. However, for the case of the lower left section, the flaring probability indicates the minimum at the first day and the maximum at the second day. This suggests that the flare forecasting based on our parameters would be best for time window 0--24 hr after the measurement of the parameters.

To make our study more useful for flare forecasting, we investigate the probability of flare occurrence as a function of each parameter of $|$$<$$\dot{H}$$>$$|$ and $<$$\Phi$$>$ for the 3 day time window, ${\tau}_{3 \textendash \mathrm{day}}$, following $\Delta t$. For this investigation, we use the flare-productive probability in the $i$th $GOES$ class, $P_{i}$, defined by
\begin{equation}
P_{i}(X)=\frac{{S_{i}}^{A}(\ge\!X)}{{S}^{T}(\ge\!X)},
\end{equation}
where $i$ represents the $GOES$ flare class and $X$ is a value of each parameter. ${S_{i}}^{A}(\ge\!\!X)$ is the number of active samples producing at least one $i$-class flare during ${\tau}_{3 \textendash \mathrm{day}}$, and ${S}^{T}(\ge\!\!X)$ is the number of the total sample in the range $[X,\infty]$. In Figure 4, we show $P_{i}$ corresponding to 14 points of the parameters for C-class as diamond symbols, M-class as triangles, and X-class as squares. Gray bars represent the number of the total $S_{T}(\ge\!\!X)$, and the dotted line denotes the range where $S_{T}(\ge\!\!X)$ is greater than 10, which maybe considered as statistically meaningful. The ratio of all the active samples in the total sample is 46$\%$, 14$\%$, and 3$\%$ for C-, M-, and X-classes, respectively. In case of $P_{i}$ as a function of $<$$\Phi$$>$, there is a fairly good linear correlation in the range of (50--570)$\times$10$^{20}$ Mx of $<$$\Phi$$>$, and the number of total samples decreases gradually. Instead, the helicity parameter shows a sharp increase in $P_{i}$ and a significant decrease in the number of total samples in the rescaled range of 0--0.15 of the parameter. $P_{C}$ as a function of $|$$<$$\dot{H}$$>$$|$, especially, quickly reaches up to $\sim$90$\%$ from 46$\%$ in the very low rescaled range, 0--0.15, of the parameter, and it retains a high value above 90$\%$ in the rest, 0.15--0.6, of the statistically meaningful range. This trend indicates that the helicity parameter can be used for differentiation between C-flare-productive and C-flare-quiet ARs. Please refer to Table 3 for further details.

For the evaluation of skill scores and success rates of the flare forecasting using the two magnetic parameters, we adopt a 2$\times$2 contingency table analysis commonly used by the meteorological and space physics communities (e.g., Fry et al. 2001, 2003). In the contingency table, there are four categories of hit, false alarm, miss, and correct null marked as $a$, $b$, $c$, and $d$ in Table 4, respectively, and defined as follows: $a$ is the number of AR samples that are predicted to produce a flare and observed with at least one flare above M-class within the 3 day time window ${\tau}_{3 \textendash \mathrm{day}}$; $b$ is the number of samples predicted to produce a flare but not observed with any flares within ${\tau}_{3 \textendash \mathrm{day}}$; $c$ is the number of samples predicted to be flare-quiet but observed with at least one flare above M-class within ${\tau}_{3 \textendash \mathrm{day}}$; and $d$ is the number of samples that were predicted to remain flare-quiet and did so within ${\tau}_{3 \textendash \mathrm{day}}$. In order to make a prediction on whether or not an AR will produce a flare, we determine a threshold of each of the two parameters as the value which makes the maximum of the Heidke skill score (HSS, hereafter) from our data sets:
\begin{equation}
\mbox{HSS} = \frac{(a+d-e)}{(N-e)},
\end{equation}
where $N$=$a$+$b$+$c$+$d$ is the total number of samples and $e$=[($a$+$c$)($a$+$b$)+($b$+$d$)($c$+$d$)]/$N$ is the number of correct forecasts by chance. HSS measures the fraction of the correct forecasts after eliminating those forecasts which would be correct due purely to random chance (Balch 2008). The positive values of HSS indicate that the forecasting performance is better than predictions by chance, and a maximum score of +1 means all correct predictions. The maximum values of HSS for $|$$<$$\dot{H}$$>$$|$ and $<$$\Phi$$>$ are 0.32 and 0.35, respectively. In addition, by using Fisher's linear discriminant analysis, we find a threshold considering both the parameters. However, the maximum HSS is 0.34 which is similar to that of each parameter. This is understandable because magnetic helicity and flux are dependent on each other somehow. For an additional assessment of the forecasting, we also consider the following quantities (Balch 2008; McKenna-Lawlor et al. 2006):
\begin{equation}
\mbox{POD} = \frac{a}{(a+c)},
\end{equation}
\begin{equation}
\mbox{FAR} = \frac{b}{(a+b)},
\end{equation}
\begin{equation}
\mbox{TCC} = \frac{a}{(a+c)} + \frac{d}{(b+d)} - 1,
\end{equation}
where POD is the probability of detection, FAR is the false alarm rate, and TCC is the true skill score used to evaluate the flaring and non-flaring accuracy. PODs are 56\% and 39\%, FARs are 64\% and 51\%, and TCCs are 39\% and 32\% for $|$$<$$\dot{H}$$>$$|$ and $<$$\Phi$$>$, respectively. Please see, in Table 4, the details of the contingency tables for $|$$<$$\dot{H}$$>$$|$ and $<$$\Phi$$>$, and the combination of $|$$<$$\dot{H}$$>$$|$ and $<$$\Phi$$>$.

Finally, in Figure 5, we present long-term (a few days) variations of the magnetic helicity calculated for 8 ARs which have the flare indexes greater than 100. We plot the magnetic helicity accumulation (cross symbols) together with the $GOES$ soft X-ray light curve (dotted line) and unsigned magnetic flux (diamond symbols) as a function of time. In the 8 ARs, there are 19 major flares with a $GOES$ peak flux greater than M5.0, and they are marked with the ID numbers of 1--19 in Figure 5. For the AR 10696 and 10720, we had already examined the helicity evolution pattern over a span of several days around the times of X-class flares which occurred in those regions in the previous paper (see Park et al. 2008). In that paper, we concluded that each of major flares was preceded by a significant helicity accumulation, 10$^{42}$--10$^{43}$ Mx$^2$ over a period of half to a few days. Another finding was that the helicity accumulates at a nearly constant rate, (4.5--48)$\times$10$^{40}$ Mx$^2$ hr$^{-1}$, and then becomes nearly constant before the flares for 4 out of 11 events. We checked these tendencies for other ARs in this study and found the followings. First, there was always a significant helicity accumulation of (3--45)$\times$10$^{42}$ Mx$^2$ before all the 19 major flares with a phase of monotonically increasing helicity over $\sim$0.5--2 days. In principle, an increase of magnetic helicity can be achieved without a flux emergence and it is frequently shown during post-flare periods for some of ARs in Figure 5. However, the increasing helicity phase before the flares always accompanied the increasing phase of magnetic flux except AR 10652, and this might be an observational result supporting the MHD simulation studies (e.g., Fan \& Gibson 2004) which show that the emergence of twisted flux ropes into pre-existing overlying field plays a critical role to produce major flares. Second, of the 19 flares, four flares (1, 5, 7, and 12) occurred when helicity injection rate becomes slow or almost zero after the significant helicity accumulation with fast injection rate. These flares are the additional examples for the almost constant helicity phase before a major flare reported by Park et al. (2008). In addition to the above two phases of helicity injection, AR 9236 and 10720 seem to have an abnormal helicity evolution pattern before the major flares compared to the monotonically increasing pattern with one sign of helicity shown in the other ARs. A remarkable feature for both ARs is that the eight major flares (2, 3, 4, 13, 14, 15, 16, and 17) occurred during the period when the helicity injection rate started to reverse its sign so that the helicity starts to accumulate with opposite sign.

\section{SUMMARY AND DISCUSSION}
We have investigated the time variations of $\dot{H}$ and $\Phi$ in 378 solar ARs and compared the two average parameters, $|$$<$$\dot{H}$$>$$|$ and $<$$\Phi$$>$, with $F_{idx}$. Although there is a large amount of scatter in the data samples, we found a moderate correlation between the parameters and $F_{idx}$. The larger $F_{idx}$ an AR has, the larger values of $\dot{H}$ and $\Phi$ it presents. To improve the correlation, we have defined a new parameter as an equally weighted linear combination of the two rescaled parameters (0.5 of each). The logarithmic-scale CC of $F_{idx}$ versus the new parameter increased slightly to 0.47. It is not surprising because $|$$<$$\dot{H}$$>$$|$ is well correlated with $<$$\Phi$$>$ as shown in Figure 3. Moreover, by considering 48 and 72 hr profiles of $\dot{H}$ and $\Phi$ for calculation of the two average parameters, we have executed the same correlation study between the two parameters and the next-day flare index. We found that the longer the period we use for average, the worse the correlation will be, especially in the case of $|$$<$$\dot{H}$$>$$|$ (CC=0.38 and 0.21 for 48 and 76 hr periods, respectively). This might be because we do not consider the flaring history before or during the measurement period of the parameters, but we compare the parameters with $F_{idx}$ calculated only for the following day of the measurement period.

We understand that no matter which method is used, the correlation between the parameters and flare index is not high. This is an intrinsic problem for flare forecasting as the occurrence of a flare depends not only on the amount of magnetic energy built up in an AR, but also on how it is triggered. For example, if new flux-rope emergence is the driver of flares (Schrijver 2009), or if a flare is exactly a result of a small and localized (quite possibly unobservable) perturbation affecting the whole system like self-organized criticality dynamics (Bak et al. 1987; B\'{e}langer et al. 2007), then it is not feasible to carry out prediction of flare onset time and magnitude by using present-day parameters derived from photospheric magnetic field observations. More specifically for our case, helicity accumulation might be a necessary, but not sufficient condition for flares. Perhaps a triggering mechanism is necessary even a magnetic system has enough non-potentiality to power a flare (so-called metastable state). This idea agrees with the study that a number of X-class flares occurred during the phase of almost constant helicity after the phase of 2--3 days of monotonically increasing helicity (Park et al. 2008).

Interestingly, contrary to the expectation that magnetic helicity injection is more closely related to flare productivity than to magnetic flux, our result shows that the correlation between $|$$<$$\dot{H}$$>$$|$ and $F_{idx}$ is not stronger than that between $<$$\Phi$$>$ and $F_{idx}$. The logarithmic-scale CCs of $F_{idx}$ versus $|$$<$$\dot{H}$$>$$|$ and $<$$\Phi$$>$ are 0.42 and 0.43, respectively. If only the flaring groups with non-zero flare index are considered, then $|$$<$$\dot{H}$$>$$|$ is not better than $<$$\Phi$$>$ in predicting how strong the flares will be. This might be due to the fact that we simply use the 1 day average of $\dot{H}$ in the entire AR for comparison with $F_{idx}$ without more specifically characterizing the temporal and spatial evolution of helicity in the AR related to a flaring condition. Magnetic helicity, however, is useful in predicting whether an AR will produce flares or not. Note that predicting the occurrence of flares is different from predicting the strength of flares. By examining more careful studies such as the helicity injection difference between flare-productive and flare-quiet ARs, the flare-productive probability as a function of $|$$<$$\dot{H}$$>$$|$, and the temporal evolution of helicity in major flare-producing ARs, we have found that magnetic helicity injection has some interesting features related to flares as follows.

\begin{enumerate}
\item For 91 AR samples in the range (3--5)$\times$10$^{22}$ Mx of large $<$$\Phi$$>$ the $flaring$ group has $|$$<$$\dot{H}$$>$$|$ about twice greater than that of the $non$-$flaring$ group. On the other hand, 118 AR samples of large $|$$<$$\dot{H}$$>$$|$ do not show the significant difference in $<$$\Phi$$>$ between the $flaring$ and $non$-$flaring$ groups.
\item The helicity parameter $|$$<$$\dot{H}$$>$$|$ demonstrates a rapid increase of $P_{i}$ compared to that of $<$$\Phi$$>$ in the rescaled range of 0--0.15 of the parameter. $P_{C}$($|$$<$$\dot{H}$$>$$|$), especially, quickly reaches up to $\sim$90$\%$ from 46$\%$ in the very low rescaled range, 0--0.15, of the parameter, and it retains a high value above 90$\%$ in the rest, 0.15--0.6, of the statistically meaningful range.
\item Helicity of (3--45)$\times$10$^{42}$ Mx$^2$ accumulates significantly and consistently over 0.5--2 days for all the 19 major flares under investigation. More specifically, following the significant amount of long-term helicity accumulation with fast injection rate, 4 and 8 flares occurred when helicity injection rate starts to become slow (sometimes almost zero) and reverse its sign, respectively.
\end{enumerate}

Based on these results, the magnetic helicity can be used for the improvement of flare forecasting. First of all, when an AR has large $<$$\Phi$$>$, we may better determine whether or not it will produce a flare by considering $|$$<$$\dot{H}$$>$$|$ of the AR. Second, the helicity parameter $|$$<$$\dot{H}$$>$$|$ would allow us to establish a better defined cutoff between C-flare-productive and C-flare-quiet ARs than $<$$\Phi$$>$ if we take into account a sharp increase of $P_{i}$ in the very low rescaled range of the parameter. Third, an early warning sign of flare occurrence could be based on tracking of a phase of monotonically increasing helicity because there is always a significant amount of helicity accumulation a few days before major flares. We might also make an urgent warning sign when helicity injection rate becomes very slow or the opposite sign of helicity starts to be injected after the significant helicity accumulation phase. The sign reversal of the magnetic helicity may support the numerical simulation model for solar flare onset proposed by Kusano et al. (2003b) in which they showed that magnetic reconnection quickly grows in the site of the helicity annihilation with different signs. Some observations of helicity inversion, similar to our result, were also reported around the time of flare onset (Kusano et al. 2003a; Yokoyama et al. 2003; Wang et al. 2004b). For more practical and advanced flare forecasting, we may think how to consider the past history of flare occurrence in an AR under investigation and combine the helicity parameter with others with different weighting coefficients. Besides that it would be required to better characterize not only the time history of helicity injection but also its spatial distribution inside ARs.

Finally, our study may lead to some physical understanding of flare on-set. For example, why do only some of the samples with the large helicity injection produce major flares, but not for all? Is a significant amount of helicity accumulation necessary or sufficient conditions for flares? We believe that the study of magnetic helicity in a coronal volume of an AR will help us to better explain physically for these questions and understand pre-flare conditions and energy storage process of flares in more detail. Our future work to calculate the coronal helicity by using three-dimensional non-linear force-free magnetic field extrapolations is therefore in progress.

\acknowledgments
We are grateful to the referee for helpful comments. This work was supported by NASA grants NNX08-AQ90G and NNX08-BA22G, and the Korea Research Foundation Grant funded by the Korean Government (KRF-2008-220-C00022). We have made use of NASA's Astro-physics Data System Abstract Service. We thank the NOAA Space Environment Center for free access to SXI data. The authors are grateful to the $SOHO$/MDI team for making the data available to use. $SOHO$ is a joint mission of ESA and NASA. MDI is funded through NASA's Solar and Heliospheric Physics program.

\clearpage

%============== BEGIN TABLES ==============

\begin{deluxetable}{lccccc}
\tabletypesize{\scriptsize}
\tablewidth{0pt}
\tablecaption{Statistical Properties of the Two Magnetic Parameters}
\tablehead{
\colhead{$X$} & \colhead{$X_{\mathrm{min}}$} & \colhead{$X_{\mathrm{max}}$} & \colhead{$X_{\mathrm{med}}\tablenotemark{a}$} & \colhead{$X_{\mathrm{avg}}\tablenotemark{b}$} & \colhead{$X_{\mathrm{stv}}\tablenotemark{c}$}}
\startdata
$|$$<$$\dot{H}$$>$$|$ (10$^{40}\,$Mx$^{2}\,$hr$^{-1}$)  & 0.011 &  94.6 & 2.15 & 7.42 & 14.9\\
$<$$\Phi$$>$ (10$^{20}\,$Mx) & 51.5 & 1180 & 274 & 320 & 206
\enddata
\tablenotetext{a}{The median value of $X$}
\tablenotetext{b}{The average value of $X$}
\tablenotetext{c}{The standard deviation of $X$}
\end{deluxetable}

\begin{table}[ht]
\scriptsize
\caption{Comparison of the Two Magnetic Parameters for $Flaring$ Groups and $Non$-$flaring$ Groups}
\vspace{25pt}
\centering
\renewcommand{\arraystretch}{1.5}
\begin{tabular}{lcccc}
\hline\hline
 & $Flaring$ Group 1 & $Non$-$flaring$ Group 1 & $Flaring$ Group 2 & $Non$-$flaring$ Group 2  \\ [0.5ex]
Sample Number & 56 & 62 & 44 & 47  \\ [0.5ex]
\hline
$|$$<$$\dot{H}$$>$$|$ (10$^{40}\,$Mx$^{2}\,$hr$^{-1}$) & 4.68 & 4.73 & 6.85 & 3.48  \\
$<$$\Phi$$>$ (10$^{20}\,$Mx) & 368 & 297 & 385 & 377  \\ [0.5ex]
\hline
\end{tabular}
\end{table}

\begin{table}[ht]
\scriptsize
\caption{Number of Active Regions Producing at Least One Flare in the $i$th $GOES$ Class During ${\tau}_{3 \textendash \mathrm{day}}$ as a Function of Magnetic Parameters}
\vspace{25pt}
\centering
\renewcommand{\arraystretch}{1.5}
\begin{tabular}{lcccccccccc}
\hline\hline
 $GOES$ Class & \multicolumn{10}{c}{Absolute average helicity injection rate  $|$$<$$\dot{H}$$>$$|$ (10$^{40}\,$Mx$^{2}\,$hr$^{-1}$)}\\ [0.5ex]
 & $\ge$0.1 & $\ge$7.4 & $\ge$14.6 & $\ge$21.9 & $\ge$29.2  & $\ge$36.5  & $\ge$43.7 & $\ge$51.0  & $\ge$58.3 & $\ge$65.5 \\ [0.5ex]
\hline
C-class & 174(378) & 93(116) & 48(54) & 33(35) & 25(26) & 17(18) & 15(16) & 10(11) & 9(10) & 8(9) \\
M-class & 54(378) & 38(116) & 21(54) & 14(35) & 12(26) & 10(18) & 9(16) & 6(11) & 6(10) & 5(9) \\
X-class & 10(378) & 9(116) & 7(54) & 6(35) & 6(26) & 4(18) & 4(16) & 3(11) & 3(10) & 3(9) \\ [0.5ex]
\hline
 $GOES$ Class & \multicolumn{10}{c}{Average unsigned magnetic flux  $<$$\Phi$$>$ (10$^{20}\,$Mx)}\\ [0.5ex]
 & $\ge$52 & $\ge$138 & $\ge$225 & $\ge$312 & $\ge$398 & $\ge$485 & $\ge$572 & $\ge$658 & $\ge$745 & $\ge$832\\ [0.5ex]
\hline
C-class & 174(378) & 163(304) & 139(221) & 112(157) & 87(110) & 60(68) & 38(43) & 27(30) & 18(18) & 11(11) \\
M-class & 54(378) & 51(304) & 45(221) & 40(157) & 34(110) & 24(68) & 21(43) & 14(30) & 10(18) & 6(11)\\
X-class & 10(378) & 10(304) & 10(221) & 9(157) & 8(110) & 7(68) & 6(43) & 5(30) & 3(18) & 1(11) \\
\hline
\end{tabular}
\tablecomments{Total number of samples are marked by parenthesis.}
\end{table}

\begin{table}[ht]
\scriptsize
\caption{Contingency Table for Evaluating the Ability of the Flare Prediction by the Two magnetic parameters}
\vspace{25pt}
\centering
\renewcommand{\arraystretch}{1.5}
\begin{tabular}{cp{0.45in}p{0.45in}p{0.45in}p{0.45in}p{0.45in}p{0.45in}p{0.45in}p{0.45in}p{0.45in}}
\cline{5-10}
& & \multicolumn{2}{c}{} & \multicolumn{2}{|c|}{$|$$<$$\dot{H}$$>$$|$} & \multicolumn{2}{|c|}{$<$$\Phi$$>$} & \multicolumn{2}{|c|}{$|$$<$$\dot{H}$$>$$|$ \& $<$$\Phi$$>$}\\ \cline{3-10}
& & \multicolumn{2}{|c|}{Forecast} & \multicolumn{2}{|c|}{Forecast} & \multicolumn{2}{|c|}{Forecast} & \multicolumn{2}{|c|}{Forecast}\\ \cline{3-10}
& & \multicolumn{1}{|c|}{Yes} & \multicolumn{1}{|c|}{No} & \multicolumn{1}{|c|}{Yes} & \multicolumn{1}{|c|}{No} & \multicolumn{1}{|c|}{Yes} & \multicolumn{1}{|c|}{No} & \multicolumn{1}{|c|}{Yes} & \multicolumn{1}{|c|}{No}\\ \cline{1-10}
\multicolumn{1}{|c|}{\multirow{2}{*}{Observation}} &
\multicolumn{1}{|c|}{Yes} & \multicolumn{1}{|c|}{a} & \multicolumn{1}{|c|}{c} & \multicolumn{1}{|c|}{30} & \multicolumn{1}{|c|}{24} & \multicolumn{1}{|c|}{21} & \multicolumn{1}{|c|}{33} & \multicolumn{1}{|c|}{22} & \multicolumn{1}{|c|}{32}\\\cline{2-10}
\multicolumn{1}{|c|}{} &
\multicolumn{1}{|c|}{No} & \multicolumn{1}{|c|}{b} & \multicolumn{1}{|c|}{d} & \multicolumn{1}{|c|}{53} & \multicolumn{1}{|c|}{271} & \multicolumn{1}{|c|}{22} & \multicolumn{1}{|c|}{302} & \multicolumn{1}{|c|}{26} & \multicolumn{1}{|c|}{298}\\\cline{1-10}
\multicolumn{2}{|c|}{Total} & \multicolumn{1}{|c|}{a+b} & \multicolumn{1}{|c|}{c+d} & \multicolumn{1}{|c|}{83} & \multicolumn{1}{|c|}{295} & \multicolumn{1}{|c|}{43} & \multicolumn{1}{|c|}{335} & \multicolumn{1}{|c|}{48} & \multicolumn{1}{|c|}{330}\\
\hline
& & & & & & & & &
\end{tabular}
\end{table}

%============== END TABLES ==============

\clearpage

%============== BEGIN FIGURES ==============

\begin{figure}
\begin{center}
\includegraphics[scale=0.85]{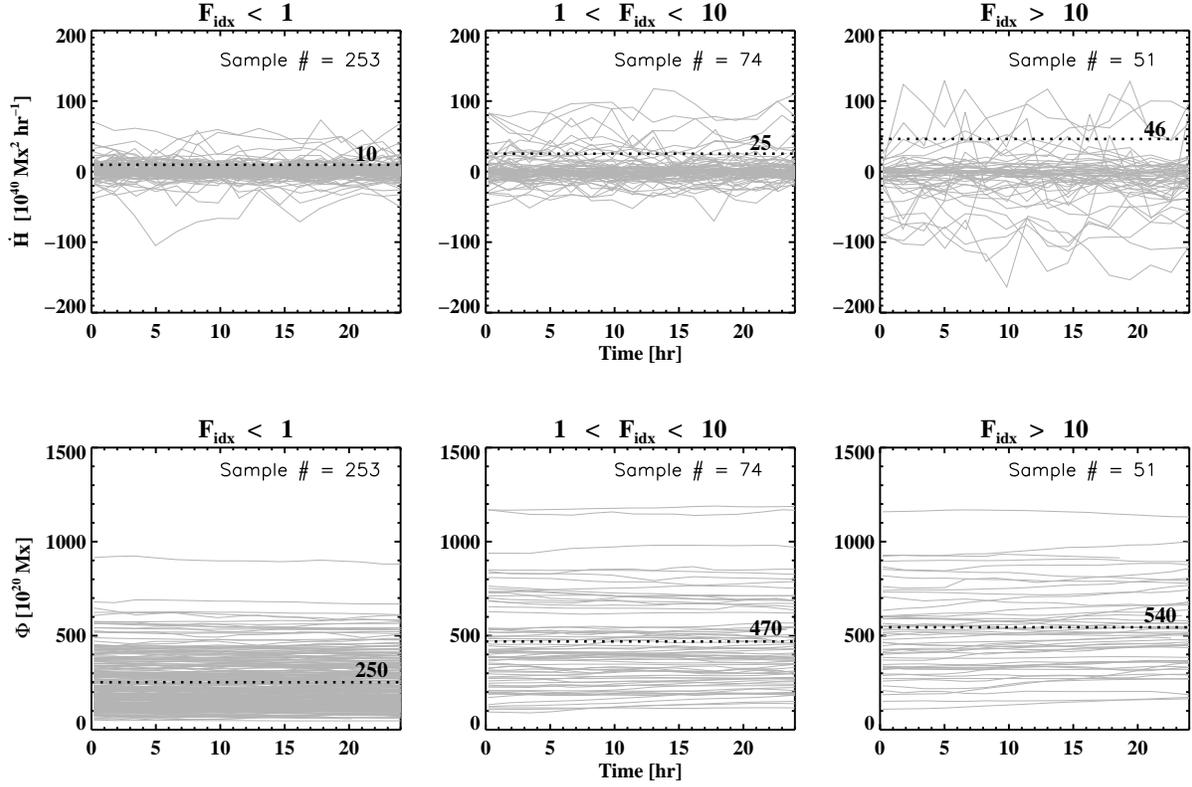}
\caption{24 hr profile of magnetic helicity injection rate, $\dot{H}$, and unsigned magnetic flux, $\Phi$. It shows three different groups classified by the ranges of flare index, $F_{idx}$, which are $F_{idx}$$<$1 (left column), 1$\le$$F_{idx}$$<$10 (middle column), and $F_{idx}$$\ge$10 (right column). The number of samples in the three groups is specified in each panel. The average maximum values for $|\dot{H}|$ (top panels) and $\Phi$ (bottom panels) of the samples in each panel are plotted as dotted lines.  \label{fig1}}
\end{center}
\end{figure}

\begin{figure}
\begin{center}
\includegraphics[scale=0.82]{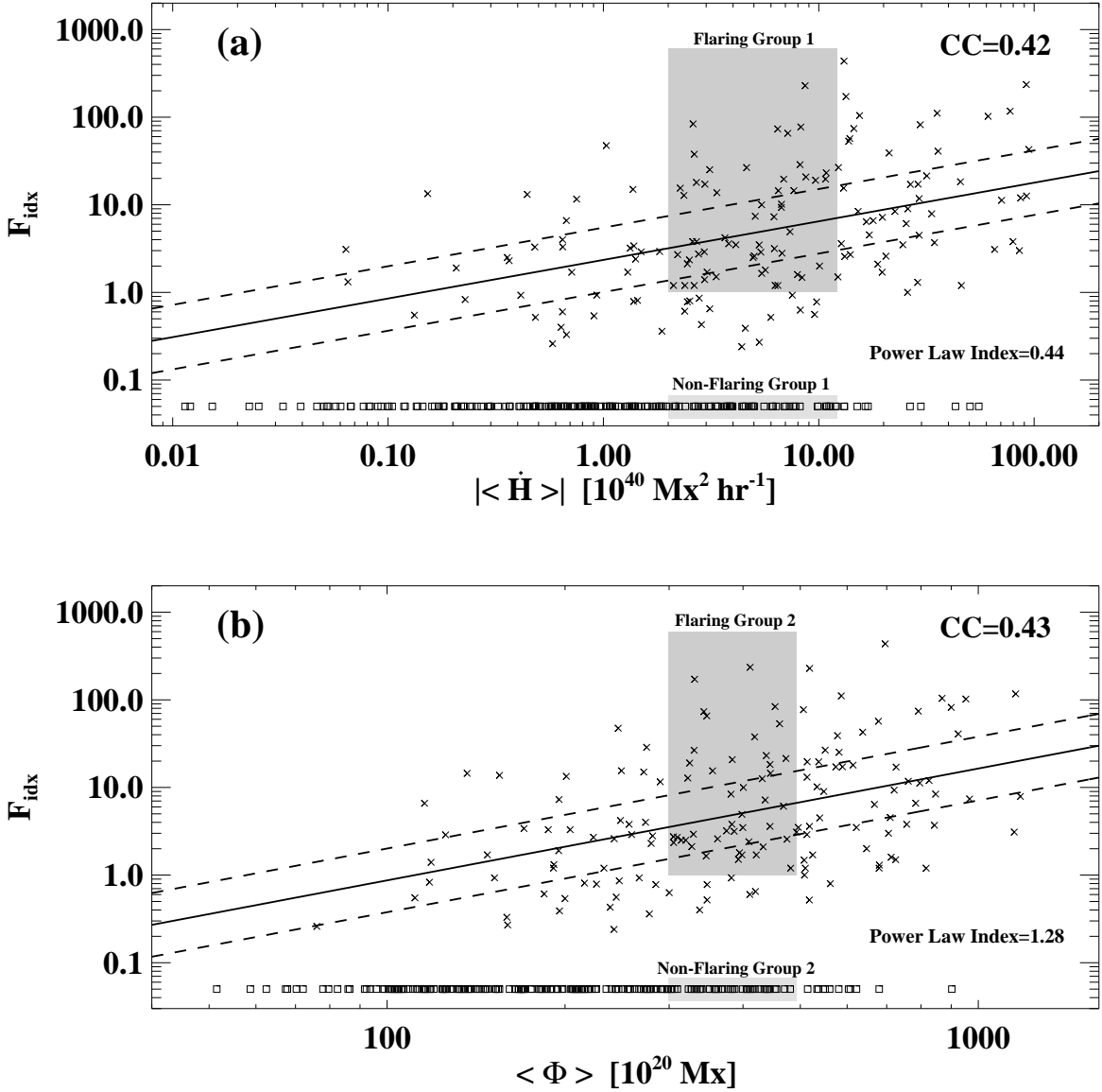}
\caption{Flare index, $F_{idx}$, vs. magnetic parameters. Correlations of $F_{idx}$ with (a) the absolute average helicity injection rate, $|$$<$$\dot{H}$$>$$|$, and (b) the average unsigned magnetic flux, $<$$\Phi$$>$, in a logarithmic scale. The solid and dotted lines show the least-squares linear fit and its standard deviation to the data points. CC is specified in each panel. The total number of samples used for the correlation studies is 153. For the other 225 samples with zero flare index, we marked them as square symbols with a small value $F_{idx}$ = 0.05 for the plotting purpose only. \label{fig2}}
\end{center}
\end{figure}

\begin{figure}
\begin{center}
\includegraphics[scale=0.7]{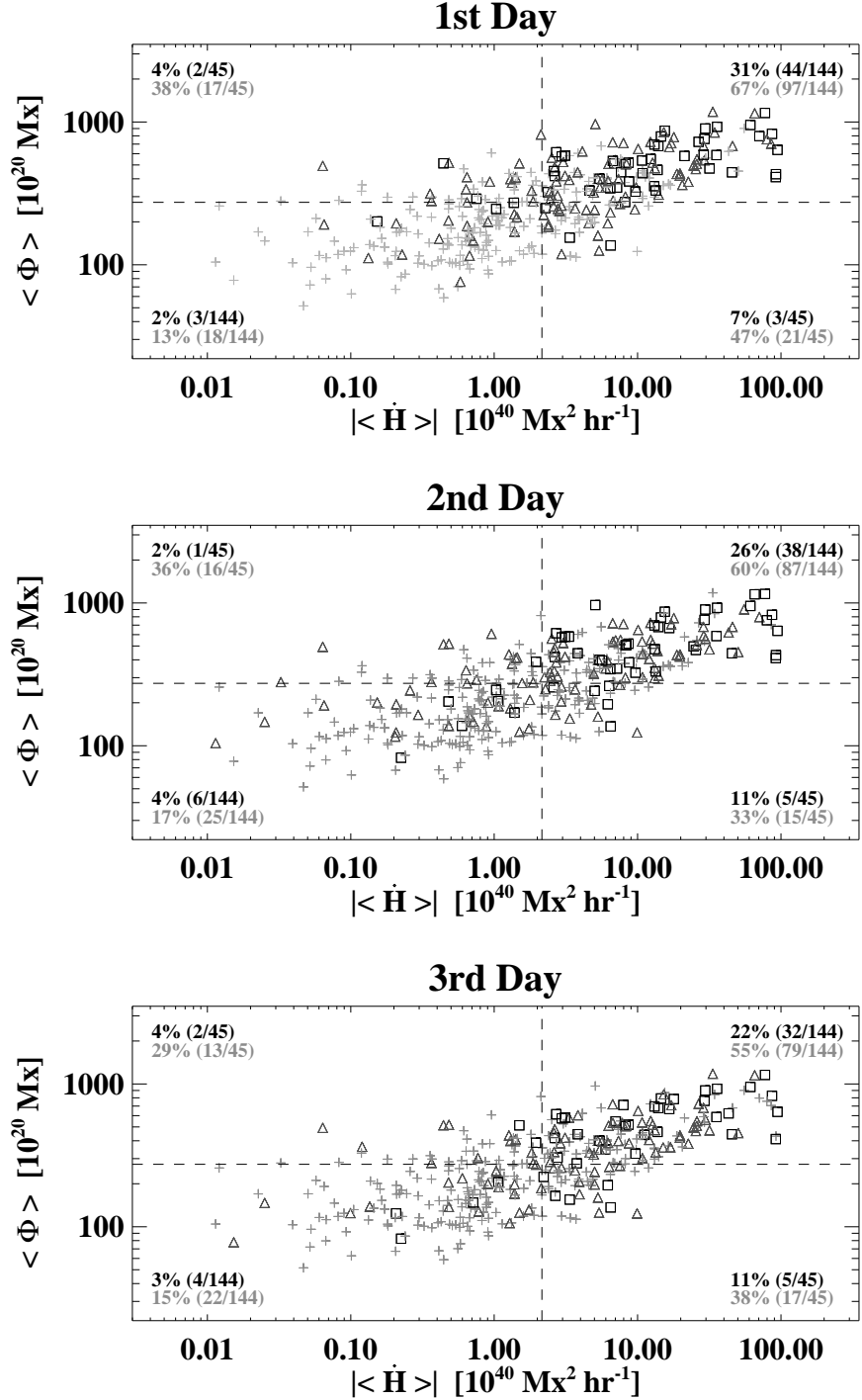}
\caption{$<$$\Phi$$>$ vs. $|$$<$$\dot{H}$$>$$|$ for the 378 samples with $F_{idx}$$<$0.1 as plus symbols, 0.1$<$$F_{idx}$$<$10 as triangle, and $F_{idx}$$>$10 as square. We calculate $F_{idx}$ for the three different time windows of the first day (top) following the 24 hr period of the parameters measurement, second day (middle), and third day (bottom). The vertical and horizontal dashed lines, in each plot, mark the median values of both the parameters and divide the domain into four sections. For the samples in each of the four sections, we calculate the probability of flare occurrence with two criteria $F_{idx}$$\ge$10 and $F_{idx}$$\ge$0.1, and marked as black and gray colored numbers in each section.  \label{fig3}}
\end{center}
\end{figure}

\begin{figure}
\begin{center}
\includegraphics[scale=0.8]{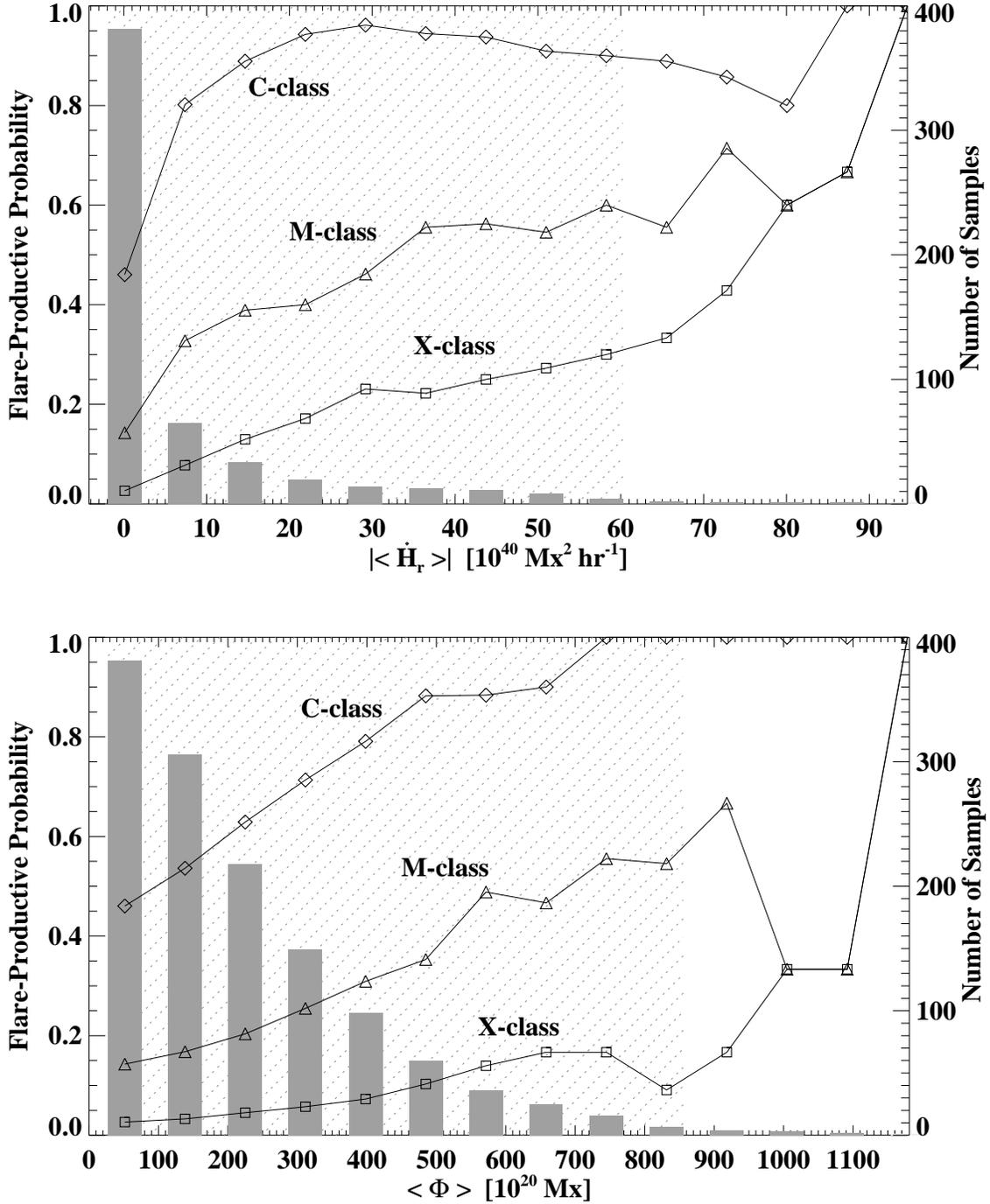}
\caption{Flare-productive probability, $P_{i}$, vs. magnetic parameters. The probabilities producing at least one C-, M-, and X-class flare during ${\tau}_{3 \textendash day}$ are shown as diamond, triangle, and square symbols, respectively. Gray bars represent the number of samples and the dotted line denotes the range where the number of the total samples is greater than 10.  \label{fig4}}
\end{center}
\end{figure}

\begin{figure}
\begin{center}
\includegraphics[scale=0.72]{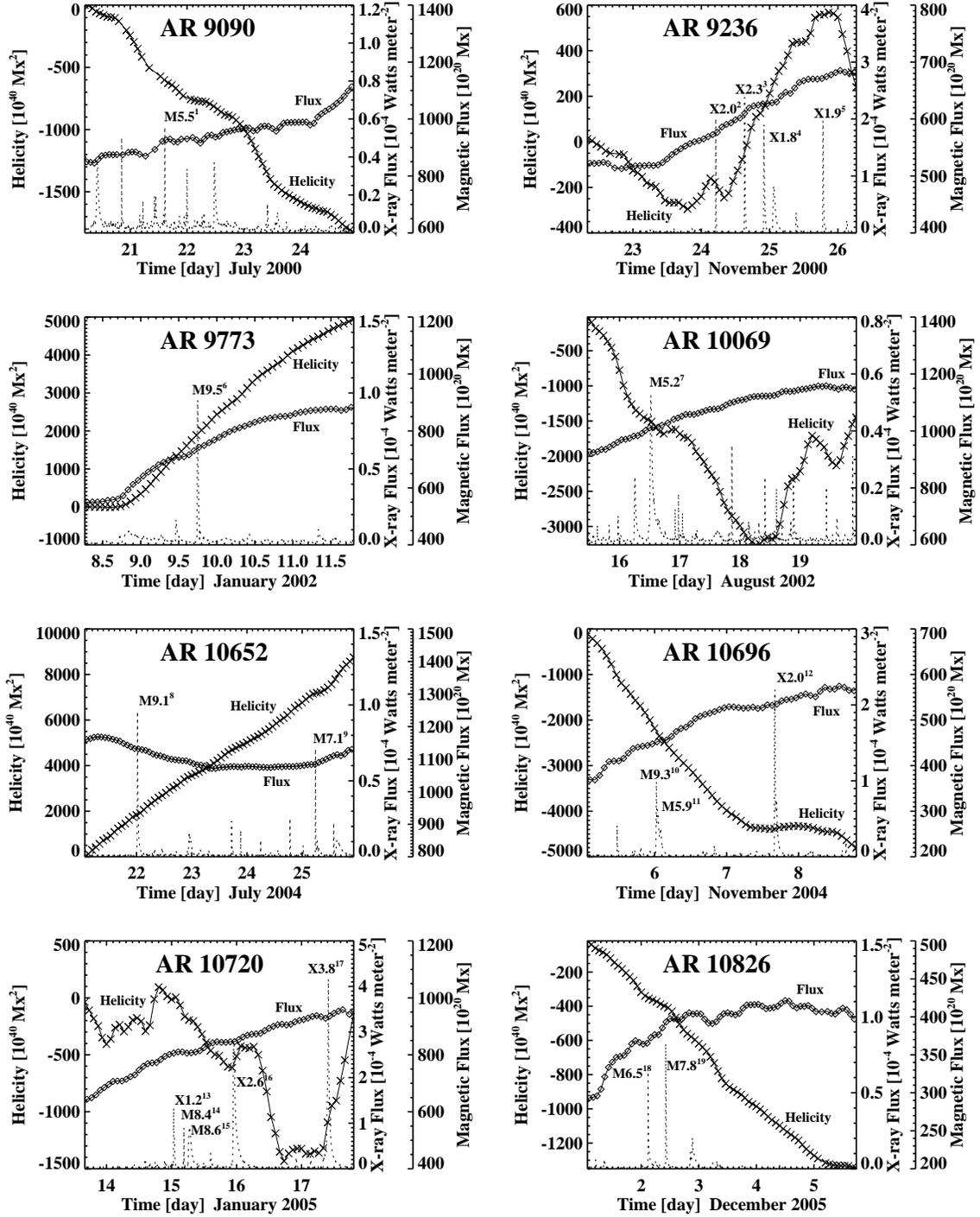}
\caption{Time variations of helicity accumulation, magnetic flux, and $GOES$ soft X-ray flux for 8 ARs which have the flare indexes greater than 100. The helicity is shown as cross symbols and the magnetic flux is shown as diamonds. The X-ray flux is shown as dotted lines and all the 19 flares above $GOES$ M5.0 level are marked with their ID numbers.  \label{fig5}}
\end{center}
\end{figure}
%============== END FIGURES ==============

\end{document}